# Experimental Investigation of Heat-generating Cheese Pallets During an International Export Cold Chain

**Dihia AGUENIHANAI*[a,b], Jean MOUREH [a], Yasmine SALEHY [a], Yvanne PAVIET-SALOMON[a], Evelyne DERENS-BERTHEAU[a]**

(a) Université Paris-Saclay, INRAE, FRISE
92761, Antony, France,
(b) CNIEL,
75009 Paris, France

*Corresponding author: dihia.aguenihanai@inrae.fr

## Abstract

This study presents an industrial experiment conducted with a dairy partner to investigate the temperature evolution of heat-generating soft cheeses within pallets during export cold chain. In the production factory, four homogeneous pallets of soft cheese were instrumented with temperature probes placed at three levels (bottom, medium, and top). Temperature was monitored over a 13 days export cold chain from France to Northern Europe. The objective was to evaluate the influence of air ventilation and pallet wrapping film on product temperature. The results clearly highlight the effect of ventilation: pallets placed in a well-ventilated area showed temperatures approximately 3°C lower than those in less ventilated zones. The study also underlines the importance of maintaining good ventilation of pallets before loading them into refrigerated transport vehicles. However, the effect of wrapping film remains unclear and does not allow definitive conclusions. This field study provides practical recommendations for better conditioning of heat-generating products in the cold chain.

## 1. Introduction

In order to preserve product quality and reduce food waste, European regulation “*(EC) N° 852/2004*“ requires industries to ensure that product temperature remains below regulatory value and that is controlled all along the cold chain. However, maintaining product temperature below the regulatory limit remains challenging, particularly for soft cheese products such as “*Camembert”* and cheese log. Indeed, soft cheese is associated with heat generation due to its microbiological activity, which directly impacts its temperature (Aguenihanai et al., 2025; Derens-Bertheau et al., 2019; Pham et al., 2021).

Moreover, the supply chain from production facilities to consumers can be complex. Products pass through multiple logistical links, including pre-cooling, storage, transportation, warehousing, retail display, and domestic refrigeration. Each link is characterized by specific operating conditions (Loisel et al., 2021), including air temperature and airflow heterogeneities. In addition, low air velocities combined with product heat generation can enhance natural convection, thereby increasing temperature heterogeneity within pallets (Aguenihanai et al., 2024). Within a given logistical link, heterogeneous airflow distribution (well-ventilated and poorly ventilated zones) may lead to the coexistence of different convective regimes. For instance, in refrigerated transport vehicles, Moureh and Flick (2004) experimentally and numerically observed high ventilation heterogeneity between the front and rear zones, where airflow stagnation and low air velocities occurred.

Beyond airflow distribution, several additional parameters may influence product temperature, such as packaging design and shape (Han et al., 2015; Nasser Eddine et al., 2022), pallet orientation (Pham et al., 2019a, 2019b), etc. In industrial practice, pallet wrapping films are widely used to ensure mechanical stability during handling and transportation. However, the impact of pallet wrapping on the thermal behaviour of heat-generating products has not been investigated in the literature.

The objective of this study is therefore (1) to better understand the impact of successive cold chain links on the temperature evolution of heat-generating products and (2) to assess the influence of ventilation conditions by comparing well-ventilated and less-ventilated zones, as well as the effect of pallet wrapping film. To this end, a 13-day logistical temperature monitoring campaign was conducted on four pallets of cheese logs from the production factory in France to Northern Europe. This field study provides practical recommendations for improving the conditioning of heat-generating products throughout the cold chain.

## 2. Materials and method

### 2.1. Design of the studied pallets

The studied pallets are homogeneous, i.e. they contain one type of product. As illustrated in Figure 1, each pallet is composed of 18 layers (levels), each containing 16 boxes measuring 30 cm in length, 15 cm in wide and 6 cm in height.
Each box contains six soft cheese legs « *Bûchette* » with a length of L = 12.5 cm, a diameter D = 4.5 cm and an individual mass of 180 g. So one pallet contains a total of 1728 cheeses per pallet. The porosity in each box is 56%, defined as the ratio between the void volume available for airflow and the total box volume.
To ensure pallet stability, five pallet interlayers are placed as follows: under the 1$^{st}$ level, between the 1$^{st}$ and 2$^{nd}$ level, 7$^{th}$ and 8$^{th}$ level, 13$^{th}$ and 14$^{th}$ level, and 16$^{th}$ and 17$^{th}$ level.

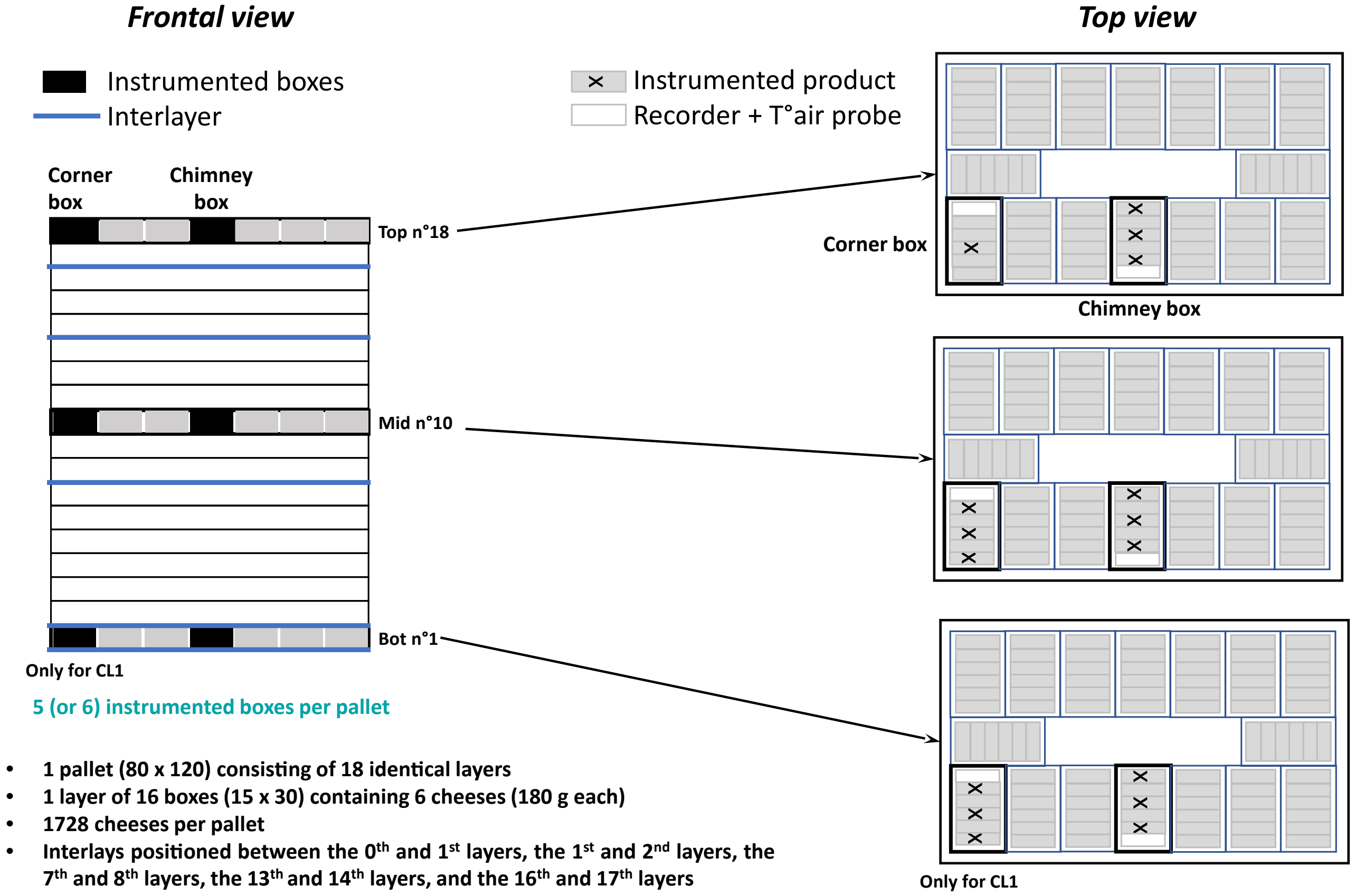

*Figure 1: Position of product and boxes instrumented within the pallet.*

## 2.2. Temperature measurements

During this field experiment, both product and air temperatures were monitored on four pallets along a supply chain, from the production facility in France to a distribution platform in Northern Europe.

Temperature measurements were carried out using Testo 176-T2 and Testo 176-T4 data loggers (Figure 2), equipped with two or four external probes, respectively. All temperature recorders were configured with an acquisition time step of 2 minutes, regardless of the type of recorder.

The pallet was instrumented as follows (see Figure 1):

- Three layers per pallet were instrumented: the bottom layer (layer 1), the middle layer (layer 10), and the top layer (layer 18).
- For the top and middle layers, two boxes were instrumented: one located at the pallet corner and one located at the center, facing the chimney zone. For the bottom layer, a single box located in the central position facing the chimney was instrumented. This configuration corresponds to five instrumented boxes per pallet, except for one pallet for which six boxes were instrumented (both corner and chimney boxes for the bottom layer).

- In each instrumented box, cheese products were equipped with temperature probes positioned at the product core, located 6.5 cm from the extremity of the cheese log. In addition, the air temperature was measured at the air gap with a probe placed above the recorder. For boxes equipped with Testo 176-T4 recorder (with four probes), three product probes and one air probe are used. For the corner box located in the top layer, a Testo 176-T2 recorder equipped with two probes (one product probe and one air probe) was used.

> **Important note:**
>
> To ensure comparison between the bottom, middle, and top pallet levels across the four pallets, the comparative analysis was performed on centrally located boxes facing the chimney zone. Corner-positioned boxes were not systematically instrumented at all three levels for each pallet, as they are often more influenced by outside air and therefore often cooler.

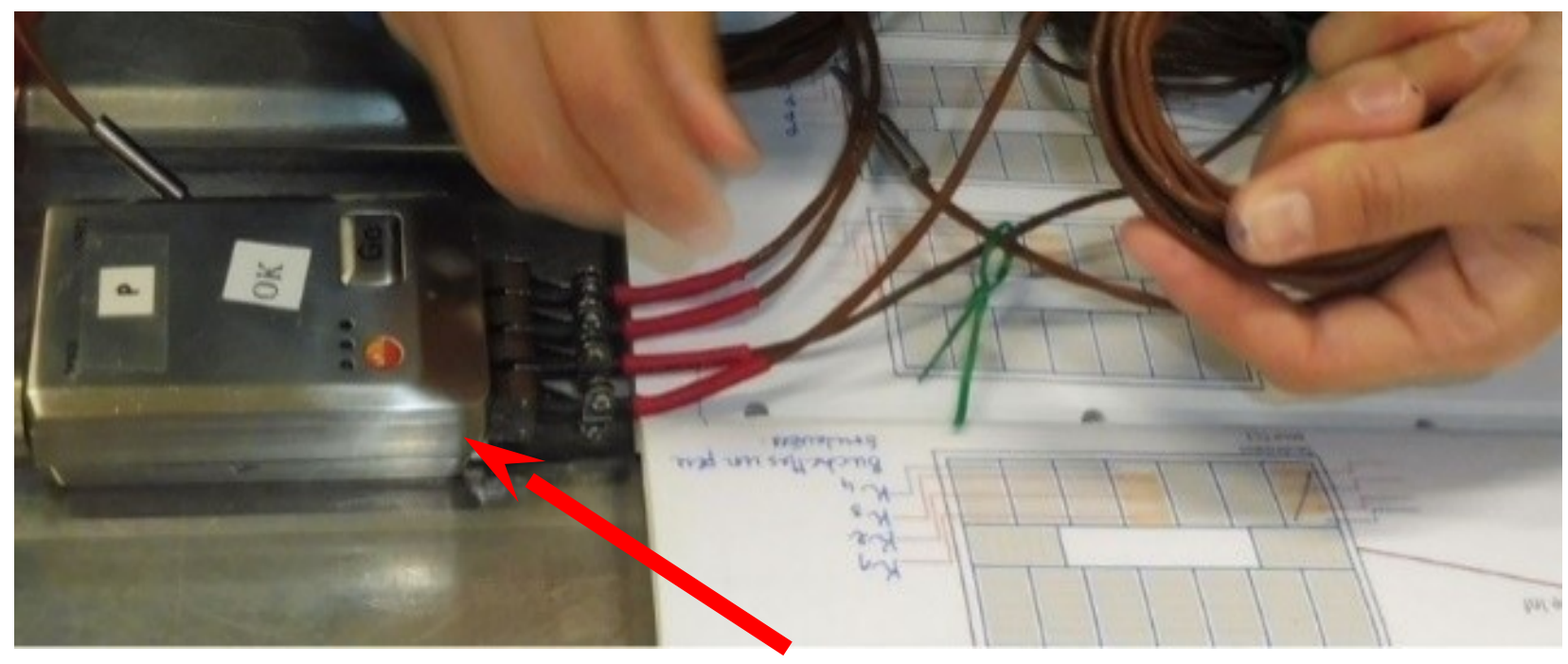

**Recorder with 4 external probes**

*Figure 2: Example of the used temperature recorders with 4 external probes.*

### 2.3. Pallet configuration and supply cold chain

Among the four instrumented pallets, two were wrapped less than the industrial practice. These pallets are referred to as *less-wrapped pallets* (MF), while pallets wrapped according to the usual industrial procedure are referred to as *conventional pallets* (CL). The MF pallets were characterized by reduced wrapping, particularly on the upper layers. However, since the conventional configuration is already slightly wrapped, the less wrapped palette (MF) shows minor differences compared to CL.

In addition, based on prior air velocity measurements conducted in the storage platform, two ventilation zones were identified: a well-ventilated zone (**Zone 1**) and a less-ventilated zone (**Zone 2**). The well-ventilated zone is located at level 4 of the storage platform, directly facing the refrigeration unit, ensuring direct contact with the cold air flow. In contrast, the less well-ventilated zone is located further away from the air supply and at a lower level, where the instrumented pallets are surrounded by other pallets, which is expected to restrict air circulation. The selection of two ventilation zones in the refrigerated vehicles: zone 1 (well ventilated) and zone 2 (less well-ventilated) is based on studies of Moureh et al. (2002, 2009) and Moureh and Flick (2005) which showed that the cold air supplied by the refrigeration unit detaches from the ceiling and generates an upstream zone characterised by high air velocities, while lower air velocities and reduced cooling efficiency occur towards the rear of the vehicle.

Conventional (CL) and less-wrapped (MF) pallets were placed in both the well-ventilated zone (blue) and the less well-ventilated zone (orange). Table 1 presents the four pallets according to their wrapping configuration and their position within the ventilation zones throughout the supply chain.

*Table 1: Identification of pallets wrapped differently in well-ventilated and less well-ventilated areas.*

| | **Zone 1 well ventilated** | **Zone 2 less well-ventilated** |
|---|---|---|
| **Conventional Pallet CL** | **CL1** | **CL2** |
| **Less-wrapped pallet MF** | **MF1** | **MF2** |

Figure 3 illustrates the pallet positioning within the storage platform and the successive stages of the cold chain. After instrumentation at the production facility, the pallets were pre-cooled for approximately 18 hours during on-site storage, then transported over a short distance (approximately 2 hours) to the company's storage platform, where they were stored for five days. The pallets were subsequently placed in the loading area for two days before being transported over three days to the distribution platform in Northern Europe, where the instrumentation recorders were removed.

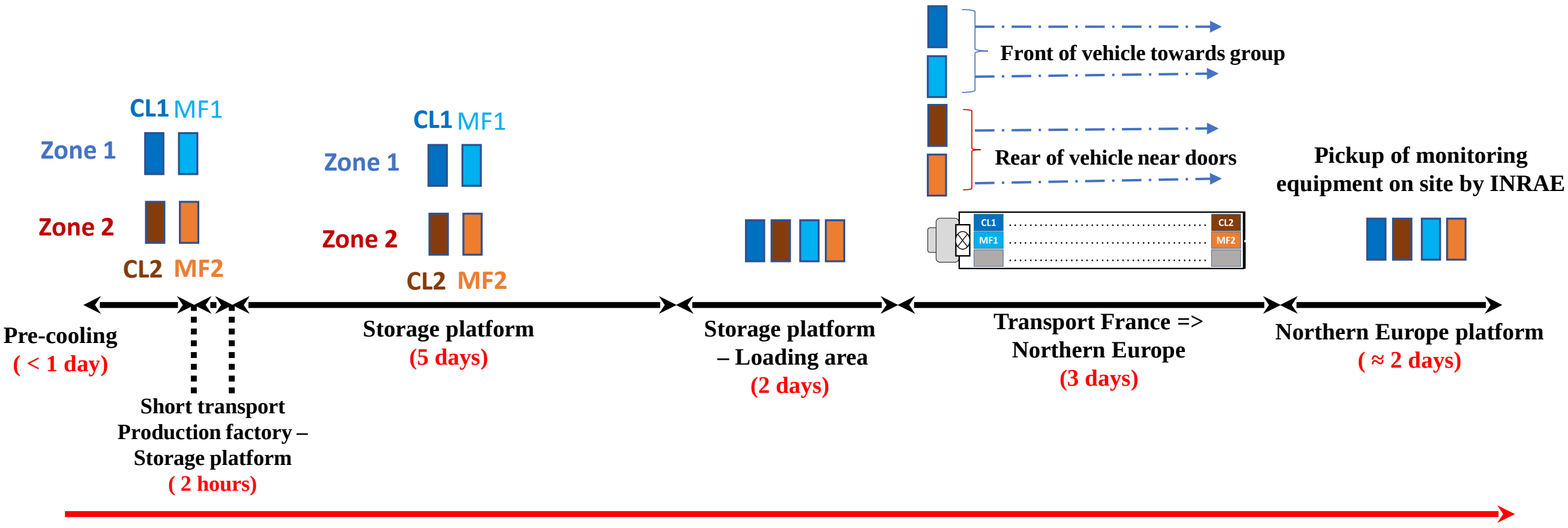

*Figure 3: Supply chain process for the four instrumented pallets.*

As shown in Figure 3, one conventional pallet (CL) and one less-wrapped pallet (MF) were placed in each ventilation zone throughout the supply chain, including at the production site, during storage, and during transportation. This experimental design allows the following comparisons:

- **Effect of pallet wrapping**: assessed by comparing MF and CL pallets placed within the same ventilation zone, i.e. CL1 versus MF1 in Zone 1.
- **Effect of ventilation conditions**: assessed by comparing pallets with the same wrapping configuration placed in different ventilation zones, i.e. CL1 versus CL2 for conventional pallets.

## 3. Results and discussion

Within each instrumented box, three product temperatures were recorded. During stabilized phases of the cold chain, these temperatures were found to be very close to each other, with differences generally below 1 °C. Therefore, an average product temperature was calculated for each box and used in the following analyses to ensure a consistent comparison between pallets and logistical stages.

The air temperature presented corresponds to the measurement taken in the box at the upper pallet level, as it is considered representative of the local air temperature and less directly influenced by product temperature. This air temperature is represented using dashed lines in the figures.

For all comparisons presented in this section, only the average product temperature measured in the centrally located box facing the chimney zone is shown for each pallet level (bottom, middle, and top). To facilitate visual comparison, temperature curves corresponding to pallets located in the well-ventilated zone (CL1 and MF1) are displayed using cold colors (blue, green, purple), while those corresponding to pallets located in the less-ventilated zone (CL2 and MF2) are displayed using warm colors (red, yellow, orange).

### 3.1. Effect of ventilation

The effect of ventilation was assessed by comparing pallets CL placed in different ventilation zones, CL1 versus CL2 (Figure 4).

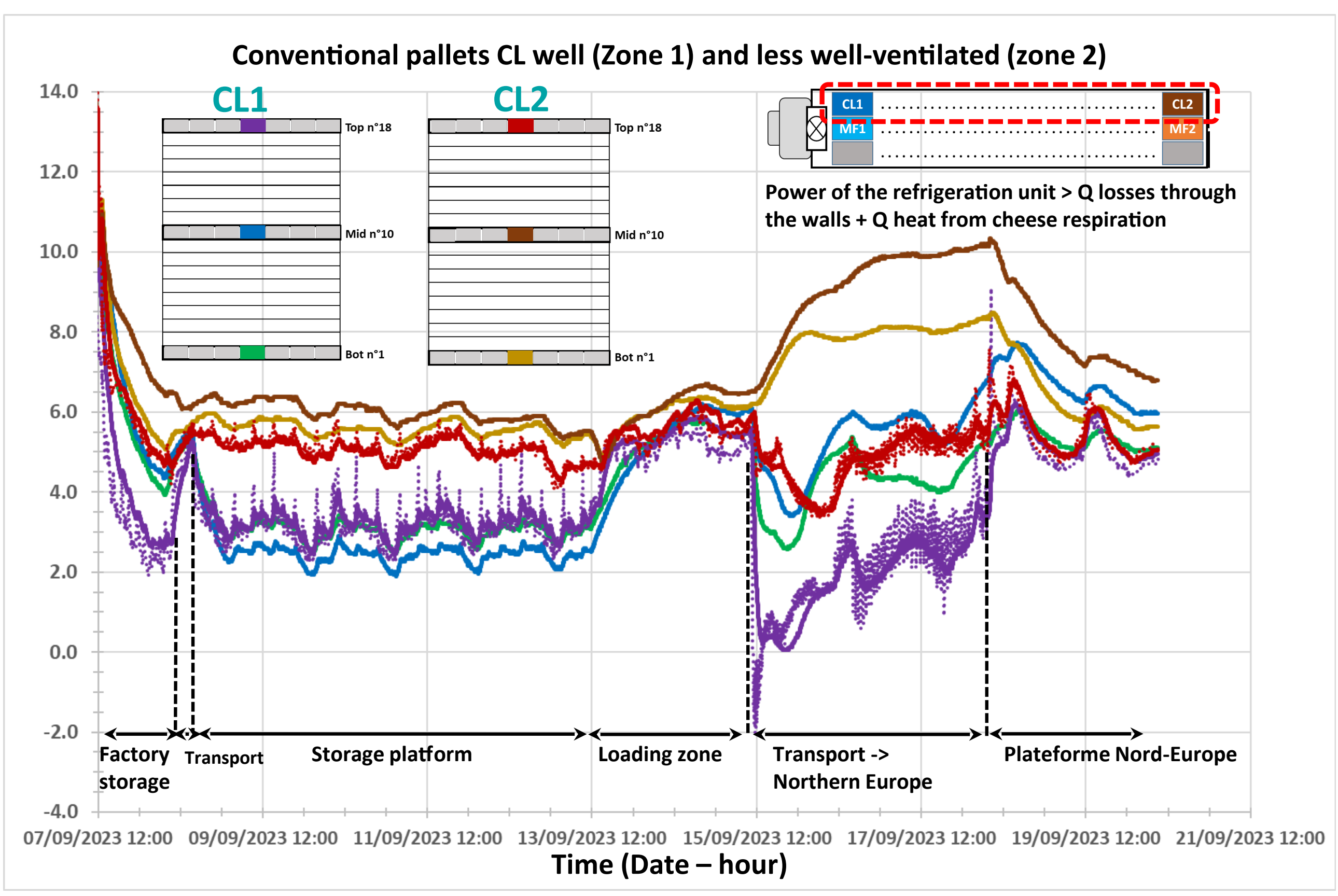

*Figure 4: Comparison of two conventionally filmed pallets placed in different ventilated areas zones. PS: Time axis is expressed as: day/month/year hour.*

**Figure 4** compares two conventionally wrapped pallets (CL1 and CL2) located in well-ventilated and less-ventilated zones, respectively. During the storage phase on the platform, a temperature difference of approximately 3 °C was observed, with average product temperatures close to 3 °C for CL1 and around 6 °C for CL2 (less well-ventilated zone).

A systematic increase in temperature was observed for all pallets during the waiting period before loading into the refrigerated vehicle. The temperature of the products reached approximately 6 to 7°C, regardless of the configuration of the pallets or their cooling history. This stage appears to be particularly critical, as it combines low air speeds (area located well downstream of the air blower) and ambient temperatures close to the regulatory value (6°C for soft cheese products). As a result, the cooling benefits achieved during 5 days of storage are rapidly lost. These observations highlight the importance of improving ventilation in this area of the platform or/and reducing the waiting time in the loading zone to approximately a maximum of 12 hours.

During transport, the CL2 pallet, located near the vehicle doors in the less-ventilated zone, reached product temperatures of up to 10 °C, whereas the CL1 pallet, positioned near the refrigeration unit, remained below approximately 6 °C. This temperature increase can be attributed to the refrigeration unit power being insufficient to fully compensate for the heat gains through the vehicle walls and the internal heat generation of the cheeses.

In well-ventilated areas of the cold chain, higher air velocities and lower air temperatures promote more effective heat removal from the products. In contrast, in less ventilated areas, lower air velocities combined with air warming in contact with upstream products generating heat result in significantly higher product temperatures. This temperature heterogeneity is similar to that already observed by Aguenihanai et al. (2025, 2024) and Pham et al. (2021).

In addition, within individual pallets, large vertical temperature gradients were observed, with the highest temperatures systematically measured at the middle pallet level. This level is less exposed to the main airflow and therefore less impacted by the ambient environment; it appears to be the most representative of the overall pallet thermal behaviour.

### 3.2. Effect of pallet wrapping

The effect of pallet wrapping was evaluated by comparing pallets placed in the well-ventilated ventilation zone: CL1 versus MF1 (Figure 5).

To improve readability within the same figure, temperature curves corresponding to conventional pallets (CL) are displayed using darker shades than those corresponding to less-wrapped pallets (MF).

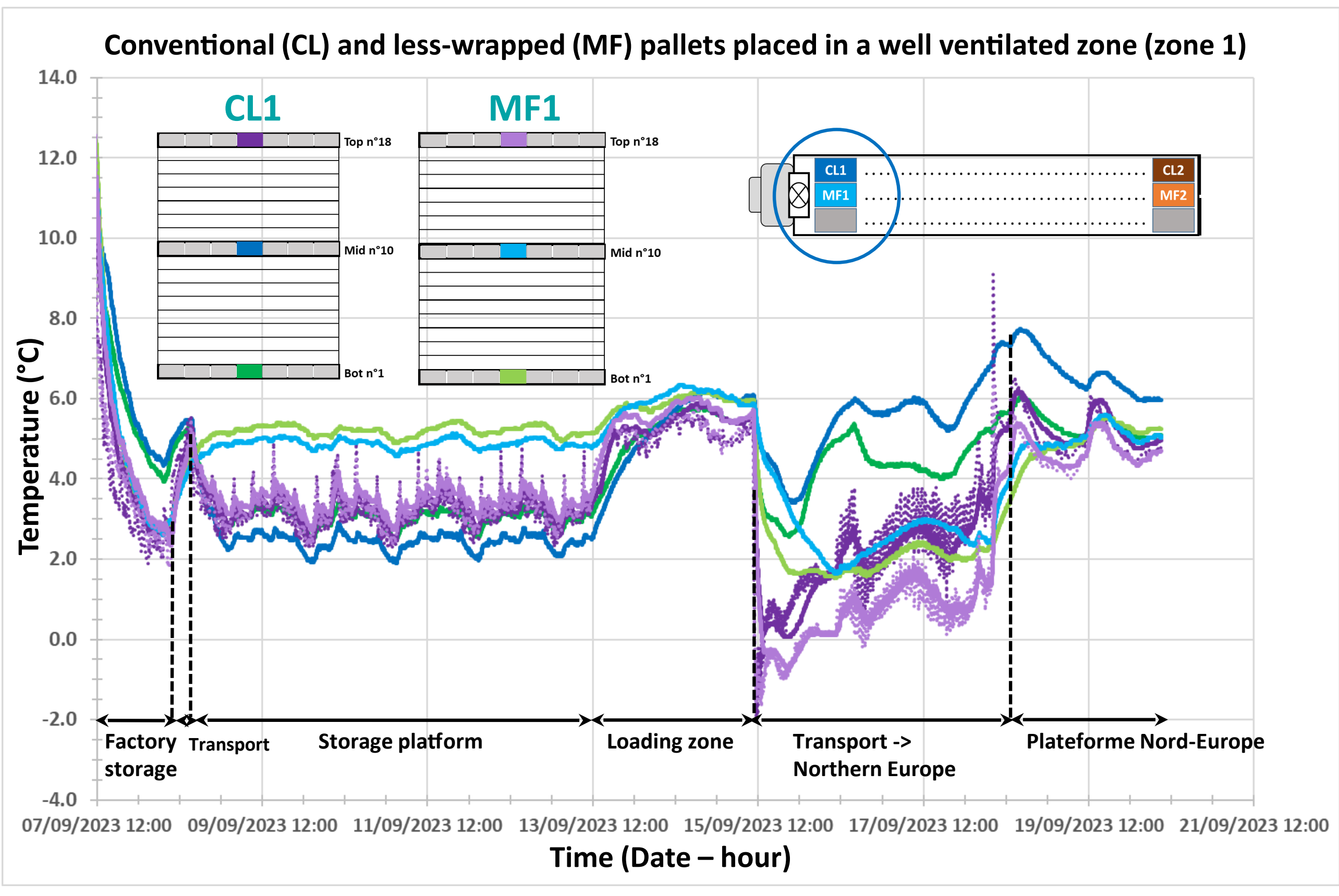

*Figure 5: Comparison of two pallets, one wrapped conventionally (CL) and the other less wrapped (MF), placed in a well-ventilated area. PS: Time axis is expressed as day/month/year hour.*

During the five-day storage period on the platform, pallet CL exhibited lower product temperatures than pallet MF (Figure 5). This difference is particularly pronounced at the middle and bottom pallet levels. This observation may be explained by the position of the pallets: CL and MF pallets are side by side, but the CL pallet is closer to the blown air. A slightly better airflow around and within CL pallets, combined with lower local air temperatures, contributed to lower product temperature compared to MF pallets. In addition, progressive air warming along the airflow direction inside the storage platform further increased this difference between the two pallets.

However, during the transport phase, an opposite trend was observed: MF pallets showed lower product temperatures than CL pallets. This behaviour may be explained by a reduced barrier to airflow due to less wrapping, allowing improved air penetration inside the pallet. In addition, MF pallets were positioned closer to the center of the vehicle and facing both the inflow and return airflows, which may also contribute to enhanced cooling. Furthermore, it is difficult to distinguish between these two effects, particularly since MF pallets are slightly less wrapped than conventional (CL) pallets.

This slight difference may not be sufficient to evaluate the effect of wrapping on pallet temperature under industrial conditions.

## 4. Conclusions

The main findings of this study can be summarized as follows:

- In the storage platform, clear ventilation heterogeneities were identified based on product temperature measurements. Pallets located near the evaporator air supply (well-ventilated zone) showed significantly lower product temperatures than pallets located in the middle of the aisle (less-ventilated zone), approximately 3 °C.
- In order to preserve the cooling benefits achieved during storage, waiting time in the loading area before transport should be kept to a minimum. Based on the observed temperature trend, a maximum waiting time of approximately 12 hours is recommended.
- During refrigerated transport, temperatures reaching 10°C were observed for pallets placed at the rear of the vehicle, near the doors (less ventilated area), while pallets located at the front, in the well-ventilated area, remained at a temperature below approximately 6.5°C. This temperature increase indicates that the refrigeration unit power was insufficient to fully compensate for the heat gains through the vehicle walls and the internal heat generation of the cheeses.

This study clearly demonstrates the dominant role of ventilation on product temperature, with lower temperature levels in well-ventilated zones compared to less-ventilated zones. The effect of pallet wrapping was less pronounced. Nevertheless, pallet wrapping is expected to limit airflow and reduce convective heat transfer within the pallet, potentially leading to higher product temperatures. For mechanical stability, the use of pallet corners rather than wrapping the pallets could be considered.

Overall, this analysis of the different links shows that the cold chain cannot be considered thermally uniform between successive links. Particular links, such as the loading area in a platform and the refrigerated transport, play an important role in the evolution of temperature heterogeneities and thermal drift. These observations highlight the need to consider the entire cold chain, taking into account not only local operating conditions, but also the cumulative effects of successive logistical stages on heat-generating products.

This work underlines the impact of ventilation and pallet configuration under real industrial conditions, characterized by heterogeneous air velocities and temperatures, and realistic logistical constraints, unlike most laboratory-based studies.

## Acknowledgements

The authors would like to express their special thanks to the dairy industrial staff who opened their doors and collaborated with us during this entire experiment. The authors acknowledge and thank the French Dairy Interbranch Organization (CNIEL) and the National French Association of Research and Technology (ANRT) for the technical and financial support that they have provided. The authors thank the CNIEL project coordinator, Fanny Tenenhaus-Aziza, for her outstanding support. The authors thank the research unit's technical team: Elyamin Dahmana, Seydina Ndoye for their technical help during the experiment.